\documentclass[aps,twocolumn,floats,preprintnumbers,prd,showpacs,nofootinbib,floatfix]{revtex4}

\usepackage{amsmath}
\usepackage{graphicx}
\usepackage{array}
\usepackage{multirow}
\usepackage{color}

\begin{document}

\preprint{UCI-HEP-TR-2014-07}

\title{Searching for Lepton Flavor Violation at a Future High Energy $e^+ e^-$ Collider}

\author{Brandon Murakami\vspace*{0.1cm}}
\affiliation{murakami@cal.berkeley.edu}

\author{Tim M.P. Tait}
\affiliation{Department of Physics and Astronomy,
University of California, Irvine, CA 92697, USA\\ ttait@uci.edu\vspace*{0.1cm}}

\date{\today}

\begin{abstract}

We consider theories where lepton flavor is violated, in particular concentrating on the four fermion operator consisting of three electrons and a tau.  Strong
constraints are available from existing searches for $\tau \rightarrow eee$, requiring the scale of the contact interaction to be $\lesssim 1 / (9~{\rm TeV})^2$.
We reexamine this type of physics, assuming that the particles responsible are heavy (with masses $\gtrsim$~TeV) such that a contact interaction
description continues to be applicable at the energies for a future $e^+ e^-$ collider.  We find that the process $e^+ e^- \rightarrow e \tau$ can be a very
sensitive probe of this kind of physics (even for very conservative assumptions
about the detector performance), already improving upon the tau decay bounds to $\lesssim 1 / (11~{\rm TeV})^2$ at collider energy
$\sqrt{s} = 500$~GeV, or reaching $\lesssim 1 / (35~{\rm TeV})^2$ for $\sqrt{s} = 3$~TeV.
Even stronger bounds are possible at $e^- e^-$ colliders in the same energy range.
\end{abstract}

\maketitle

\section{Introduction}

With the discovery of the Higgs boson \cite{Aad:2012tfa,Chatrchyan:2012ufa} at the Large Hadron Collider, the Standard Model (SM) of
particle physics is complete and self-consistent.  Nonetheless, it still misses important ingredients necessary to describe Nature, and
there is much about it that is mysterious.  Among its chief mysteries is the explanation for the tiny neutrino masses inferred from their
propagation over large distances.  The SM itself contains accidental flavor symmetries for the leptons which would prevent such masses
from occurring.  Their presence is suggestive of physics beyond the Standard Model which violates these symmetries, inspiring
a rich experimental program of searches for lepton flavor violating (LFV) 
processes \cite{Marciano:2008zz}.

Sources of LFV physics can be classified according to the masses of the particles mediating the flavor violation.  If these particles
are light, they must be very weakly interacting (or more precisely, the flavor-violating part of the interaction must be very small) in order to remain
consistent with null results for experimental searches for LFV.  This is typically the case considered when
flavor violation is included in the minimal supersymmetric standard model via a mismatch between the mass bases of the charged leptons and
their superpartners \cite{Lee:1984kr,Lee:1984tn,Hall:1985dx,Borzumati:1986qx,Gabbiani:1988rb,Hisano:1995nq,Hisano:1995cp,Gabbiani:1996hi,Hisano:1998wn,Hisano:1998fj,Cannoni:2003my,Deppisch:2003wt,Masiero:2004js,Calibbi:2006nq,Feng:2009bs,Feng:2009bd,Koike:2010xr,Abada:2012re}.
In such a situation, the most promising probes of LFV are typically through a combination of low energy searches for rare processes and collider searches aimed
at directly observing the particles mediating the interaction themselves.  Collider searches for this case, modulo the need for enough energy to produce the
mediators on-shell, do not typically benefit from extremely high energies.  Instead, the reach typically depends on the rate of production and how distinctive the
resulting signals are compared to the relevant backgrounds.

If the mediating particles are much heavier, the flavor violating couplings can be much larger.  If the masses are large compared to the energies accessible to
the collider, the underlying details become less important and their effects can be described by local operators in an effective field theory.  In this regime,
high energy reactions have the advantage that the rate for the LFV processes grow with energy, and thus the highest energy colliders have the largest
lever arm to probe lepton flavor violation.

In this article, we examine LFV in this high mass regime, focusing on operators that take the form of $\overline{e} e \overline{e} \tau$ contact interactions.
While this is but one choice of many combinations, it serves to illustrate the capabilities of a future high energy $e^+ e^-$ or $e^- e^-$ collider to tell
us something about the nature of flavor in the lepton sector.  The choice of an interaction which combines electrons with taus (as opposed e.g. to a choice like $\overline{e} e \overline{e} \mu$) is motivated by the
fact that bounds on this particular type of interaction are among the weakest in the charged lepton sector, with the current bounds provided by
searches for the decay $\tau \rightarrow e e e$ by BELLE \cite{Hayasaka:2010np}.   We study the four fermion operators (instead of the flavor-changing magnetic/electric dipole moments) because the four fermion operators are dimension six and thus grow with energy, resulting in a relatively improved lever arm
compared to the search for rare decay processes.

As we will see below, our choice also illustrates how a future high energy lepton collider
can fill in the gaps that are more difficult for low energy searches, and allows us to explore the requirements for detectors given the challenge
of a final state containing a tau lepton.  There is some overlap in our work with Ref.~\cite{Ferreira:2006dg}; we devise a different search strategy
which leads to improved prospects for an observation.
It is already well understood that a future high energy lepton collider can probe the couplings of the
Higgs boson \cite{Baer:2013cma,Peskin:2012we,Klute:2013cx}
and electroweak interactions of the top quark \cite{Baer:2013cma,Batra:2006iq} to exquisite precision.
As we demonstrate below, it can also dramatically improve our understanding of LFV processes.

\section{$\overline{e} e \overline{e} \tau$ Operators}

The lepton flavor violating interactions of interest consist of four fermion
operators of the form $\overline{e} \Gamma_a e \otimes \overline{e} \Gamma_b \tau$, where 
$\Gamma_{a,b}$ are elements of the set 
$\{ 1, \gamma_5, \gamma_{\mu}, \gamma_{\mu} \gamma_5, \sigma_{\mu \nu} \}$, and
any free Lorentz indices are contracted to form an invariant.  There ten Lorentz invariant combinations.
Under a Fierz transformation  \cite{Nieves:2003in}, each operator in the set transforms into a linear combination
of the same operators, implying that six of them may be re-expressed in terms of the remaining four.
We choose the remaining operators in linear combinations such that each bilinear consists of
a vector operator with a specific chirality.  As a result, the fully general
set of  $\overline{e} e \overline{e} \tau$ LFV interactions is contained in the Lagrangian density,
\begin{eqnarray}
& & V_{LL} \left[ \overline{e} \gamma^\mu P_L e \right] \left[ \overline{\tau} \gamma_\mu P_L e \right]
+ V_{RR} \left[ \overline{e} \gamma^\mu P_R e \right] \left[ \overline{\tau} \gamma_\mu P_R e \right]
\nonumber \\ & &
+ ~ V_{LR} \left[ \overline{e} \gamma^\mu P_L e \right] \left[ \overline{\tau} \gamma_\mu P_R e \right]
+ V_{RL} \left[ \overline{e} \gamma^\mu P_R e \right] \left[ \overline{\tau} \gamma_\mu P_L e \right] 
\nonumber \\ & & + {\rm ~H.~c.}
\end{eqnarray}
where the $V_{ij}$ ($i,j=L,R$) parameterize the strength of each interaction, and are complex coefficients
of mass dimension -2 representing the heavy fields that have been integrated out, leaving behind these
interactions as the least suppressed (by the energy of the reaction) remnant of their existence.
These operators are consistent with the full SU(3)$\times$SU(2)$\times$U(1) gauge symmetry of the SM.

\subsection{Constraints from $\tau \rightarrow e e e$}

There are important constraints on the magnitude of the $V_{ij}$ coming from null searches for the
decay $\tau \rightarrow e e e$.  The current best limit is provided by BELLE, which places an upper limit
on the branching ratio of \cite{Hayasaka:2010np},
\begin{eqnarray}
B \left( \tau \rightarrow e e e \right) \leq 2.7 \times 10^{-8}~.
\end{eqnarray}
In terms of the $V$'s (and neglecting $m_e \ll m_\tau$), the partial width for the flavor violating
decay is given by,
\begin{eqnarray}
\Gamma 
& = & \frac{m_\tau^5}{1536 \pi^3} \left( |V_{LR}|^2 + |V_{RL}|^2 + 2 \left(  |V_{LL}|^2 + |V_{RR}|^2 \right) \right)
~.
\nonumber \\
\end{eqnarray}
Combined with the BELLE limit, this restricts the magnitude of the $V$'s to:
\begin{eqnarray}
 \left( |V_{LR}|^2 + |V_{RL}|^2 + 2 |V_{RR}|^2 + 2|V_{LL}|^2 \right) \nonumber \\  \leq 1.63 \times 10^{-16}~{\rm GeV}^{-4} = 
 \frac{1}{(8845~{\rm GeV})^2}~.
\end{eqnarray}
Future $b$ factory experiments \cite{Abe:2010gxa,Bona:2007qt} could potentially improve on these bounds
by as much as two orders of magnitude.

OPAL \cite{Abbiendi:2001cs} and BABAR \cite{Aubert:2006uy} have also searched for the process
$e^+ e^- \rightarrow \tau e$.  Null results from those experiments can also be used to bound a combination of
the $V$'s, but these bounds are weaker than the decay bounds from BELLE.

\section{Signal Rates}

We are interested in the reactions $e^+ e^- \rightarrow e \tau$ and $e^- e^- \rightarrow e^- \tau^- $.
In the limit where the center of mass energy $\sqrt{s}$ is much larger than the tau mass, the differential
cross section for $e^+ e^- \rightarrow e \tau$ is given by,
\begin{eqnarray}
\frac{d\sigma}{dt} & = &
\frac{1}{16 \pi s^2}
\Big\{
4 \left( |V_{LL}|^2 + |V_{RR}|^2 \right) u^2 
\nonumber \\ & & ~~~~~~ + 
\left( |V_{LR}|^2 + |V_{RL}|^2 \right) \left( t^2 + s^2 \right)
\Big\}
\end{eqnarray}
where $t = (p_\tau - p_e^i)^2$ and $u = -s - t$ are the usual Mandelstam invariants.
Integrating this expression over $t$ results in the total cross section for
$e^+ e^- \rightarrow e \tau$,
\begin{eqnarray}
\sigma \left( s \right) & = &
\frac{s}{12 \pi}
\Big\{ |V_{LL}|^2 + |V_{RR}|^2 +  |V_{LR}|^2 + |V_{RL}|^2 \Big\} .~~~
\label{eq:sigma}
\end{eqnarray}

The expressions for the process
$e^- e^- \rightarrow e^- \tau^-$ is easily obtained from the previous one by crossing the initial state
$e^+$ with a final state $e^+$.  The resulting cross section is,
\begin{eqnarray}
\sigma \left( s \right)  = 
\frac{s}{16 \pi}
\left\{ 4 \left( |V_{LL}|^2 + |V_{RR}|^2 \right) +  \frac{2}{3} \left( |V_{LR}|^2 + |V_{RL}|^2 \right)
\right\} .~~~
\label{eq:sigma}
\end{eqnarray}

\section{Search Strategy and Projected Limits}

The $e^+ e^- \rightarrow \tau e$ signal events are characterized by an electron and a tau lepton, produced back-to-back in the center
of mass frame and with energies very close to $\sqrt{s} / 2$.  Our selection strategy attempts to reconstruct both electron and tau, and
require that they are close to back-to-back and with the correct energies.  While
the tau decay will necessarily produce missing momentum, the
ability to uniquely reconstruct the center of mass frame provides a very powerful discriminant allowing the signal to be filtered from the
(otherwise) overwhelming background processes.
Our cuts are not very dependent on the specifics of the detector design, and we comment
on the assumptions concerning the detector performance below.

We assume that the visible tau decay products can be identified as likely resulting from a tau decay, but our reconstruction is not
very dependent on a specific decay mode.  A tau decay in the SM will produce at least one charged particle (typically $e^\pm$, $\mu^\pm$, 
or $\pi^\pm$), one or more neutrinos, and in some cases some neutral hadrons.  

There is a fake background arising from $e^+ e^- \rightarrow q \bar{q}$,
where one of the resulting hadronic jets produces only a few pions (and thus fakes a tau candidate)
and the other fakes an electron; this background
will be negligible compared to other processes provided the probability for a jet of hadrons to fake an electron ($P_{j \rightarrow e}$)
multiplied by the rate to fake a tau ($P_{j \rightarrow \tau}$) is 
\begin{eqnarray}
P_{j \rightarrow e} \times P_{j \rightarrow \tau} & \lesssim & 10^{-2} ~.
\end{eqnarray}
Given the excellent performance anticipated for the detectors proposed at future high energy $e^+ e^-$ facilities
\cite{Behnke:2013lya,Gomez-Ceballos:2013zzn}, it is likely that performance will exceed
this requirement by at least an order of magnitude, and this fake background can be safely discarded.

There are two reducible backgrounds involving real taus and electrons.  The electroweak process
$e^+ e^- \rightarrow \tau \nu_\tau e \nu_e$ produces a real tau and a real electron and
is the dominant background for $\sqrt{s} \gtrsim 500$~GeV.  This background arises dominantly from intermediate
weak bosons produced approximately on-shell, and for energies in the range of 100 GeV to TeV, it grows
with energy as the phase space for the resonant particles increases.
In addition, the process
$e^+ e^- \rightarrow \tau^+ \tau^-$ produces electrons from the decay $\tau \rightarrow e \nu_e \nu_\tau$,
which will rarely produce an electron whose energy is $\simeq \sqrt{s} / 2$.  We simulate both of these
backgrounds using the {\tt MadGraph 5} package \cite{Alwall:2011uj}.

\subsection{Event Selection}

Uncut, these backgrounds are about a thousand times larger than the largest signal consistent with the constraints
from tau decays.  We reduce them to a manageable level by reconstructing the events as follows.
First, we identify the highest energy electron in the event, and require that it have energy
\begin{eqnarray}
E_e & \geq & \left( 1 - r \right) ~ \frac{\sqrt{s}}{2}~.
\end{eqnarray}
Since the signal typically produces electrons with $E_e = \sqrt{s}/2$, it is desirable to choose $r$ to be as small as
is feasible.  We find that even a rather modest choice of $r = 0.1$ is sufficient, which is well below the
expected energy resolution for isolated electrons \cite{Behnke:2013lya,Gomez-Ceballos:2013zzn}.  This cut is
very effective, reducing both backgrounds by ${\cal O}(10^{2})$ for $\sqrt{s} \lesssim 1$~TeV.  At higher energies,
it becomes less effective in dealing with the $e^+ e^- \rightarrow \tau \nu_\tau e \nu_e$ background, but is still 
modestly helpful.

We can further reduce the $e^+ e^- \rightarrow \tau \nu_\tau e \nu_e$ background by exploiting the fact that it
rarely results in back-to-back events.  After removing the primary electron, the remaining
visible particles would correspond to the visible tau decay products in a signal event.  The sum of their momenta
is denoted $\vec{p}_{\rm vis}$.  We can also construct the net missing momentum of the event 
(using the knowledge of the center-of-mass frame, the measured primary electron momentum, and $\vec{p}_{\rm vis}$).
In a signal event, this missing momentum (\mbox{$\not \hspace*{-0.05cm} \vec{p}$}) is entirely from the neutrino(s) produced
in the tau decay, and so under the signal hypothesis we can reconstruct the tau momentum as 
\begin{eqnarray}
\vec{p}_\tau & = & \vec{p}_{\rm vis} + \not \hspace*{-0.05cm} \vec{p}~.
\end{eqnarray}
From here, we reconstruct the center of mass frame as $\overline{s} = (p_e + p_\tau)^2$ and require,
\begin{eqnarray}
\sqrt{\overline{s}} & \geq & \left( 1 - r \right) \times \sqrt{s}
\end{eqnarray}
where for simplicity, we choose the same $r = 0.1$ as before.  It should be noted that reconstructing
$\overline{s}$ typically involves reconstructing hadrons from the tau decay; an energy resolution on the order of
$10\%$ remains a conservative choice \cite{Behnke:2013lya,Gomez-Ceballos:2013zzn}.  This cut on
$\overline{s}$ can be understood to be equivalent
to requirement that the reconstructed tau and electron are approximately back-to-back, and is defined
to be robust under the presence of visible final state radiation photons.

We note in passing that this reconstruction strategy, which takes advantage of the knowledge of the center-of-mass frame
available in an $e^+ e^-$ collider environment, and the fact that an event consisting of a single tau plus
visible particles can be completely reconstructed, could also be useful in dealing with $t \bar{t}$ events where
one top decays into a tau lepton, or lepton-flavor violating Higgs decays into $\tau e$ or $\tau \mu$.
Essentially the same analysis (with the final state $e$ replaced by a muon) should also work to
bound the process $e^+ e^- \rightarrow \mu \tau$; but this process receives contributions from a wider
set of dimension six operators, and we leave quantifying such bounds for future work.

\subsection{Initial State Radiation}

At leading order and given the conservative choices of $r$, the signal events are expected to pass these cuts
with near-perfect efficiency.  However, radiation of photons which are lost along the beam axis will lead to a slight
degradation of the signal, as such photons cannot be accounted for in $\overline{s}$.  The fraction of events
containing a photon approximately collinear with the beam and carrying a momentum fraction $x = E_\gamma / E$ of the
original beam electron (or positron) energy $E$ can be approximated \cite{Landau:1934zj,Brodsky:1971ud},
\begin{eqnarray}
f_{\gamma \leftarrow e} \left( x, E \right)
& = & \frac{\alpha}{2 \pi} ~ \log \left( \frac{E_\gamma^2}{m_e^2} \right) ~ \frac{1 +(1-x)^2}{x}~.
\end{eqnarray}
Based on this expression, we estimate that about $10\%$ of the signal events will contain a
beam-collinear photon carrying sufficient energy to result in the event failing to pass either
the $E_e$ or the $\sqrt{\overline{s}}$ cut.  At a future linear collider, beamstrahlung is expected to
be similar in size to initial state radiation; for a more detailed discussion see \cite{pandora}.

\begin{table}[t]
\begin{center}
\caption{Expected number of events for the $e \tau$ signal (assuming $V_{LL} = 1 / ({\rm 10~TeV})^2$
and $V_{RR} = V_{RL} = V_{LR} = 0$)
and background processes, assuming $1~{\rm ab}^{-1}$ of data collected at an $e^+ e^-$ collider
running at $\sqrt{s} = 250, 500, 1000,$ and $3000$ GeV, before and after the cuts described in the text.
Also indicated is the expected $95\%$ C.L. on $V_{LL}^{-1/2}$ assuming no signal is observed.}
\label{tb:events}
\begin{tabular}{ccccc}
\hline\hline \\[-0.325cm]
~Process~ & ~~~$250$~GeV~~~ & ~~~$500$~GeV~~~ & ~~~$1$~TeV~~~ & ~~~$3$~TeV~ \\ \hline \\[-0.20cm]
 & & Before Cuts & & \\
$e \tau$ Signal & $112$ &  $450$ & $1800$ & $1.6 \times 10^4$ \\
$e \tau \nu_e \nu_\tau$ & $4.6 \times 10^5$ & $5 \times 10^5$ & $6.6 \times 10^5$ & $1.2 \times 10^6$ \\
$\tau \tau$ &$6.3 \times 10^5$  & $1.5 \times 10^5$ & $3.7 \times 10^4$ & 4200 \\[0.1cm]
 & & After Cuts & & \\
$e \tau$ Signal & $101$ & $405$ & $1620$ & $1.5 \times 10^4$ \\
$e \tau \nu_e \nu_\tau$ & $9300$ & $10^4$& $5900$ & 2480 \\
$\tau \tau$ & 6590 & 1600 & 390 & 44 \\ \hline \\[-0.325cm]
$V_{95}^{-1/2}$ & 8.0~TeV & 11.7~TeV & 18.0~TeV & 34.9~TeV \\[0.1cm]
\hline\hline
\end{tabular}
\end{center}
\vspace*{-0.2in}
\end{table}

\subsection{Projected Limits}

Assuming a collected data set of $1~{\rm ab}^{-1}$, we estimate the potential $95\%$ C.L. limits on the couplings
$V_{LL}, V_{RR}, V_{RL},$ and $V_{LR}$ for future $e^+ e^-$ colliders with energies
$\sqrt{s} = 250$, $500$, $1000$, and $3000$ GeV.  As a reference value consistent with the limits from tau decays,
we set $V_{LL} = 1 / (10~{\rm TeV})^2$ and $V_{RR} = V_{RL} = V_{LR} = 0$.  From Eq.~(\ref{eq:sigma}), it is obvious
how to translate these bounds onto theories where different combinations of the couplings $V_{ij}$ are non-zero.

In Table~\ref{tb:events}, we show the events predicted for the signal process (under these assumptions) as well as the 
backgrounds, both before and after the cuts described above are applied.  After cuts, the signal to background ratio is
1:10 at the lowest considered energies, and more like 100:1 at the highest.  It would be worthwhile to explore tightening the
cut parameter $r$ at lower energies to take advantage of the very precise detectors proposed for future $e^+ e^-$ colliders,
but we leave this kind of optimization for future work once the detector design parameters have been more firmly established.
Also shown in Table~\ref{tb:events} are the projected $95\%$ C.L. limits on $V_{LL}^{-1/2}$ for each collider energy.  
At $\sqrt{s} = 250$~GeV, the projected limits are comparable but slightly worse than those currently available from tau decays.
These limits steadily improve with collider energy, reaching $\sim 35$~TeV for $\sqrt{s} = 3$~TeV.

If it were to prove feasible to tighten the cut parameter to $r = 0.05$, greater background rejection is possible, though at the cost
of larger signal losses ($\sim 30\%$) due to initial state radiation.  The net result would nonetheless be a modest gain in sensitivity up to
$V_{LL}^{-1/2} \geq \{ 9.3, 14.2, 20.3, 39.5 \}$~TeV for $\sqrt{s} = \{ 250, 500, 1000, 3000 \}$~GeV.
Further improvements could be obtained by polarizing the incoming beams, which could reduce the $e \tau \nu_e \nu_\tau$ background,
though the cost to the signal would depend on which of the $V_{ij}$ are dominant.  This feature might be best exploited
to disentangle the chiralities involved once a signal is observed and we leave detailed exploration of such refinements to future work.

\subsection{$e^- e^-$ Collider}

At a high energy $e^- e^-$ collider, the signal consists of $e^- e^- \rightarrow e^- \tau^-$.  The dominant background is
$e^- e^- \rightarrow e^- \nu_e \tau^- \overline{\nu}_\tau$, and there is no analogue to the $\tau \tau$ background.  
The same strategy employed above for $e^+ e^-$ collisions
should be effective at extracting the signal from this background.  Given the smaller backgrounds, and assuming
$1~{\rm ab}^{-1}$ of collected data and $r = 0.1$, we find projected limits 
$V_{LL}^{-1/2} \geq \{ 18.6, 21.9, 27.0, 42.2 \}$~TeV for $\sqrt{s} = \{ 250, 500, 1000, 3000 \}$~GeV.
Even at $\sqrt{s} = 250$~GeV, this is a substantial improvement over the existing constraints from
tau decays.

\section{Conclusions}

We have examined the constraints on models of $\overline{e} e \overline{e} \tau$
LFV in which the particles mediating the interaction
are heavier than a few TeV.  In this regime, the physics is captured by four contact operators, with four complex
coefficients that parameterize the details of the underlying model.  We find that a future $e^+ e^-$ or
$e^- e^-$ collider with $\sqrt{s} \gtrsim 250$~GeV is able to expand our knowledge of such interactions
well beyond the already strong constraints coming from tau decays.

There are many ways in which a high energy lepton collider would expand our knowledge of particle physics;
lepton flavor violation is one of them.

\section*{Acknowledgments}
The authors are grateful for conversations with P. Tanedo, A. Roodman,
and especially T. LeCompte for an initial question which eventually metamorphosed in this project;
and to M. Peskin for insightful comments on a draft.
The research of TMPT is supported in part by NSF
grant PHY-1316792 and by the University of California, Irvine through a Chancellor's fellowship.
The work of BM was supported in part by NSF award PHY-1068420.
BM thanks the UC Irvine Particle Theory Group for hosting him and providing fruitful discussions during various visits during 2013-14.  

\end{document}